\documentclass[twocolumn, twocolappendix]{aastex631}
\let\tablenum\relax 
\usepackage{amsfonts,amsmath,graphicx,natbib,url,hyperref,nicefrac,siunitx}
\usepackage{xcolor}

\usepackage{mathrsfs}

\usepackage{amssymb,verbatim,subfigure,paralist,soul,comment,todonotes}
\sethlcolor{green} 
\usepackage{float}
\usepackage[normalem]{ulem} 
\usepackage[dvipsnames]{xcolor}
\hypersetup{colorlinks=true, urlcolor=blue, citecolor=blue}

\begin{document}

\title{A novel method for dust evolution in protoplanetary disks and a new criterion for gap opening}

\author[0000-0001-6891-275X]{S. M. Gopika}
\author[0000-0001-7626-9629]{Paul C. Duffell}
\affiliation{Department of Physics and Astronomy, Purdue University, West Lafayette, IN 47907, USA}

\begin{abstract} We present the \textit{Grain Residence-Time Method} (GRTM), a computationally efficient particle-tracking approach for studying dust evolution and trapping in protoplanetary disks. By reconstructing steady-state dust distributions from particle residence times, this method provides a way to obtain radial dust surface density profiles without directly evolving the dust as a fluid.  We compare the GRTM with dust profiles computed via standard methods using the \texttt{FARGO3D} code, which models the dust as a pressureless fluid, and find very good agreement over a wide range of parameter space.  Finally, we investigate how dust gap formation depends on the planet-to-star mass ratio ($q$) and Stokes number ($\mathrm{St}$), through a parameter study, establishing an empirical criterion for dust gap opening. In doing so, we have discovered a new regime of gap opening for small grains with $St \lesssim 10^{-2}$ and for planet mass well below Jupiter's, which is likely very relevant for gaps seen in ALMA images.
\end{abstract} 
\section{Introduction} \label{sec:intro}

High-resolution observations with the Atacama Large Millimeter/submillimeter Array (ALMA) have revealed a rich diversity of substructures in protoplanetary disks, including concentric rings and gaps, spiral arms, and azimuthal asymmetries \citep{andrews2018,huang2018}. Despite extensive theoretical and numerical investigations, the physical origins of these structures remain uncertain, and several mechanisms have been proposed, including photoevaporative and magnetically driven disk winds, hydrodynamic and magnetohydrodynamic instabilities, and gravitational instabilities \citep{armitagereview_protoplanetarydisks,2007dullemond,andrews2020}. In particular, embedded planets have been proposed as a mechanism for producing gaps in the gas and generating pressure maxima that can trap drifting dust particles \citep{pinilla2012,zhangdsharp}. Hydrodynamic simulations coupled with radiative-transfer calculations have also been used to investigate whether planet-induced gaps and rings can reproduce observed disk emission, including the multiple ring and gap structures observed in HL Tau \citep{dong2015,jin2016}.

Gas gaps carved by planets have been extensively studied using analytical models and hydrodynamic simulations \citep{crida2006,gap_paul,duffell2020,duffelgasgap,Kanagawa2015,Kanagawa2016,fung2014}. These studies have established criteria for gas-gap opening and investigated how gap depth and width depend on the planet-to-star mass ratio and disk properties. \cite{rafikov2002} developed a semi-analytical treatment of planet-generated density waves and their role in gap formation, showing how the dissipation of these waves can modify the gas surface density. Together, these studies provide a framework for understanding how embedded planets perturb the gas disk. However, understanding the observed substructures also requires considering how dust responds to these planet-induced changes in the gas. 

Dust grains experience aerodynamic drag due to their relative motion with respect to the gas. Because the gas is partially supported against gravity by its radial pressure gradient, it generally rotates at a sub-Keplerian velocity. Dust grains therefore experience a headwind, lose angular momentum, and drift radially inward toward the central star \citep{whipple1972,weidenschilling1977}. The drift velocity depends on the aerodynamic coupling of the grains to the gas, commonly characterized by the Stokes number ($\mathrm{St}$), as well as on the local disk conditions. Consequently, dust transport can differ substantially from gas transport. Planet-induced perturbations can modify radial dust drift by altering the gas pressure distribution and creating pressure maxima. These maxima can halt the inward drift of dust grains, leading to dust accumulation outside the planet's orbit \citep{zhu2012}.  The efficiency of dust filtration across planet-induced gaps also depends on dust evolution \citep{jtlaune2020}. In particular, fragmentation can replenish the population of small grains that are able to cross a gap, allowing dust to reach the inner disk even when larger grains are efficiently filtered \citep{pfiel2025}. 
 
These results highlight the importance of considering both the aerodynamic coupling of the grains and the processes that alter their size distribution when studying dust transport.
Dust back-reaction can also influence the structures formed through planet-disk interactions. While dust grains experience aerodynamic drag from the gas, they also exert an equal and opposite drag force on the gas. This feedback can modify the gas velocity and density structure, particularly when the dust-to-gas ratio becomes sufficiently high. Two-fluid hydrodynamic simulations have shown that dust feedback can broaden dust rings outside planet-induced gaps \citep{kanagawa2018}. It can also modify the gas and dust morphology, including spiral structures, gap edges, and pressure maxima, thereby affecting dust filtration \citep{yangzhu2020}.
Dust gap formation has also been investigated in previous studies, which have explored its dependence on the planet-to-star mass ratio and the coupling between dust and gas \citep{dipi_gas_dust,dipnlaibe}. These studies have shown that a dust gap can form even when there is no gap in the gas, particularly for solids with larger aerodynamic sizes. \cite{ataiee} has also examined the conditions under which an embedded planet produces
a pressure maximum in the gas, enabling the trapping of drifting solids. However, the dependence of dust gap formation and particle trapping outside the planet's orbit on the planet-to-star mass ratio ($q$) and Stokes number ($\mathrm{St}$) remains to be systematically tested.

A variety of numerical approaches have been developed to study dust evolution in protoplanetary disks. In Eulerian multi-fluid methods, gas and dust are evolved as separate fluids on a spatial grid, with the dust typically treated as a pressureless fluid. Such approaches have been used to investigate dust dynamics and planet–disk interactions \citep{paardekooper2006}, and are implemented in numerical frameworks such as \texttt{FARGO3D} \citep{Masset2000,benitzandmasset,fargo3d} and \texttt{Athena++} \citep{athenadust}. Three-dimensional radiative two-fluid hydrodynamic simulations have also been used to investigate planet-induced dust stirring \citep{binkert2023}, as well as meridional circulation and the delivery of solids into circumplanetary regions \citep{szulagyibinkert2022}.
An alternative is to treat the gas as a fluid while representing dust grains as individual particles. This hybrid approach has been adopted in studies using particle-based dust dynamics, including \cite{plutowithdustparticle,Paardekooper_dustpart, rapid2026}. More recently,  \cite{clepper2025} combined \texttt{FARGO3D} hydrodynamic simulations with particle tracking to investigate dust transport in planet–disk systems. The choice of numerical scheme for particle integration has been studied in detail, including measurements of the impact of these numerical choices on the solution \citep{funganddhruv2019}.

Another widely used approach is two-fluid smoothed particle hydrodynamics (SPH), a mesh-free Lagrangian method in which both the gas and dust are represented by particles. In the two-fluid formulation, gas and dust are represented by separate sets of SPH particles that interact through aerodynamic drag \citep{Laibe2008, fouchet2007,ayliffe2012}. These methods have been applied to study dust settling, radial migration, and the formation of planet-induced dust gaps and pile-ups \citep{fouchet2007,dipi_gas_dust,dipnlaibe}. One-fluid formulations have also been developed for gas–dust mixtures in SPH, in which the gas and dust are treated as components of a single fluid rather than as two separately evolved fluids\citep{pricelaibe2012,pricelaibe2012b,pricelaibe2014}.


In this work, we introduce the \textit{Grain Residence-Time Method (GRTM)} for reconstructing steady-state radial dust surface-density profiles from particle residence times. Rather than directly evolving a dust-fluid density field, we integrate particle trajectories through a prescribed gas velocity field obtained from \texttt{FARGO3D} simulations and use the time spent by particles in radial bins to reconstruct the dust distribution. This provides an alternative approach to studying dust transport and trapping while retaining information about individual particle trajectories.


The GRTM is highly efficient (on the order of 10-30 minutes on a single-core laptop computer), enabling a large parameter survey of planet-to-star mass ratios ($q$) and Stokes numbers($\mathrm{St}$), to empirically measure the criteria for gap opening. We present such a parameter study which in addition to confirming the predictions of previous theoretical works, exhibits a new regime of dust gap opening at low $\mathrm{St}$ numbers ($St \lesssim 0.01$), where the dust drift rate is slow but the dust is still not completely coupled to the gas.

The remainder of this paper is organized as follows. In Section \ref{sec:methods}, we describe the gas background, particle integration, and residence-time reconstruction procedure. Section \ref{sec:code_tests} presents the numerical convergence tests and comparisons with the \texttt{FARGO3D} dust-fluid results. In Section \ref{sec:results}, we do a parameter study to examine the dependence of dust gap formation on the planet-to-star mass ratio and Stokes number. Finally, in Section \ref{sec:discussion}, we discuss the key results, limitations, and possible future extensions of the method.
\begin{figure*}[!t]
    \centering
    \includegraphics[width=\textwidth]{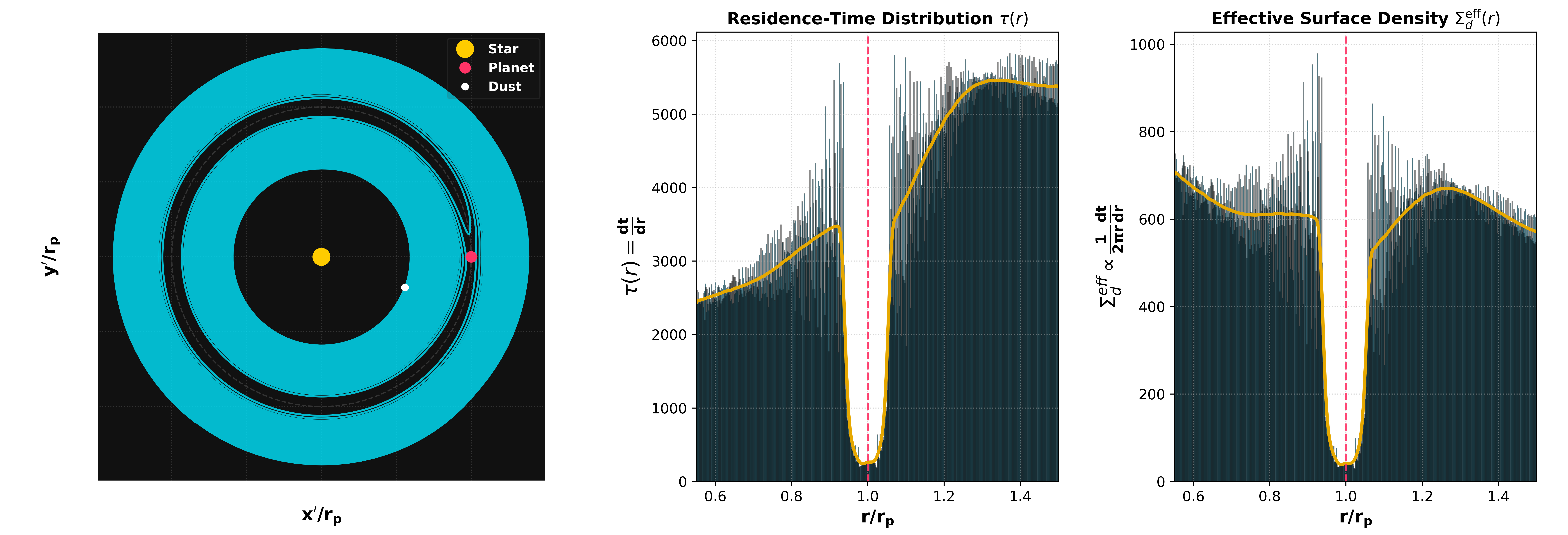}
    \caption{Demonstration of the Grain Residence- Time method for computing dust surface density. The left panel shows the particle trajectories in the 2D corotating frame, while the right panel shows the residence time ($ \frac{dt}{dr}$) and effective surface density ($\frac{1}{2 \pi r} \frac{dt}{dr}$) histograms. The red dashed line indicates the planet's position ($r = r_p$) and the yellow solid line represents the Gaussian smoothed density profile. For this demonstration, we consider $q = 3\times10^{-4}$, $\alpha = 5\times 10^{-3}$, Mach number, $\mathcal{M}=10$, Stokes number $St =0.02$ and the potential smoothening length to be $\varepsilon = 0.5H$, where $H$ is the disk scale height.}
    \label{fig:particle_method}
\end{figure*}

\section{Methods} \label{sec:methods}

In this section, we describe the Grain Residence-Time Method and its application to reconstructing radial dust surface density profiles. We first describe the prescribed gas background, followed by the particle dynamics and the reconstruction of the dust distribution from particle residence times.

\subsection{Gas background from \texttt{FARGO3D}}

\texttt{FARGO3D} \citep{fargo3d} is a numerical code developed with particular emphasis on protoplanetary disk physics and planet-disk interactions. It solves both hydrodynamic and magnetohydrodynamic equations and is parallelized using MPI, with support for both CPU and GPU architectures. In this work, we use \texttt{FARGO3D} in its hydrodynamic configuration to obtain the gas velocity field used as the prescribed background in the GRTM. We also employ its multifluid implementation, described in \cite{fargo3d}, to evolve dust as a pressureless fluid for comparison with our method.

For the GRTM, we use the gas velocity field from the final  snapshot from \texttt{FARGO3D} output at $t=4000$ orbits as a prescribed background during the particle integrations. The gas velocity is interpolated onto the particle positions using bilinear interpolation. The choice of this output is motivated by the time-evolution analysis of densities presented in Section \ref{timeseries}. The specific initial conditions and parameters adopted for the \texttt{FARGO3D} simulations are described in Section \ref{sec:code_tests}.

Although the Stokes number generally depends on local disk properties, we treat it as a constant parameter in our residence-time method and the corresponding \texttt{FARGO3D} dust-fluid simulations for simplicity and for ease of comparison.  This assumption can be easily relaxed in future studies.  

\subsection{Particle Dynamics}

Particle trajectories are integrated in an inertial frame using a fourth-order Runge–Kutta method. The equations of motion are
\begin{align}
  \frac{d\mathbf{x}_d}{dt} &= \mathbf{v}_d, \\
  \frac{d\mathbf{v}_d}{dt} &= \mathbf{a}_{\rm grav} - \frac{\Omega_K}{St}\left(\mathbf{v}_d-\mathbf{v}_g\right),
\end{align}
where $\mathbf{x}_d$ and $\mathbf{v}_d$ denote the particle position and velocity, respectively, and $\mathbf{v}_g$ is the gas velocity, interpolated from \texttt{FARGO3D} outputs. The Stokes number is defined as
\begin{equation}
\mathrm{St} = \Omega_K t_{\rm stop},
\end{equation}
where $\Omega_K$ is the local Keplerian angular frequency and $t_{\rm stop}$ is the aerodynamic stopping time, which characterizes the timescale over which drag brings the particle velocity toward the gas velocity.

We integrate $N$ independent particle trajectories with different initial azimuthal phases to test the sensitivity of the dust distribution to the particles' initial positions. The initial dust velocities are prescribed analytically ($v_{r,d} = v_{r,d}^{NSH}$ and $v_{\phi,d} = v_K$), while the subsequent
particle evolution is calculated using the gravitational forces and the gas
velocity field interpolated from \texttt{FARGO3D} outputs.

The gravitational acceleration includes contributions from the central star and the planet:
\begin{equation}
\mathbf{a}_{\rm grav} = - G M \frac{\mathbf{r}-\mathbf{r}_\star}{|\mathbf{r}-\mathbf{r}_\star|^3} - G m_p \frac{\mathbf{r}-\mathbf{r}_p}{\left(|\mathbf{r}-\mathbf{r}_p|^2+\varepsilon^2\right)^{3/2}},
\label{eq:ref_smoothening}
\end{equation}
where $G$ is the gravitational constant, $M$ and $m_p$ are the stellar and planetary masses, and $\mathbf{r}_\star$ and $\mathbf{r}_p$ are their respective positions. The parameter $\varepsilon$ is the planetary potential softening length. Although the GRTM can integrate particle trajectories using an unsoftened potential ($\varepsilon=0$), we use a finite softening length in our comparisons with \texttt{FARGO3D}, which also employs a softened planetary potential. The effect of planetary potential softening on the reconstructed dust distribution is examined in Section \ref{subsec:eps}.

\subsection{Reconstruction of dust surface density from particle residence times}

The key idea of the Grain Residence-Time Method is to infer the radial dust distribution from the time particles spend in different radial intervals. We divide the computational domain into radial bins of width $\Delta r$ and accumulate the time spent by the particles in each bin. For each particle, the residence-time distribution is defined as
\begin{equation}
\tau(r_i) = \frac{1}{\Delta r} \sum_k \Delta t_k \, \mathbb{I}_{r_i}(r_k),
\end{equation}
where $\Delta t_k$ is the time interval associated with the $k$-th time integration step of the particle, and $\mathbb{I}_{r_i}(r_k)$ is an indicator function that equals unity when the particle is in radial bin $r_i$ and zero otherwise.

In the continuous limit, the residence time per unit radial distance is related to the particle's average drift velocity as,
\begin{equation}
\tau(r) = \left|\frac{dt}{dr}\right|
= \frac{1}{|v_{r,d}(r)|},
\end{equation}
where $v_{r,d}$ is the radial dust velocity. Thus, particles moving slowly through a radial bin (remain in or repeatedly pass through a given radial bin) contribute more strongly to its residence time, whereas particles crossing it rapidly contribute less. 

For a steady radial flow with a constant dust mass flux,
\begin{equation}
\dot{M}_d = 2\pi r \, \Sigma_d(r)\, |v_{r,d}(r)|= \mathrm{constant},
\end{equation}
where $\Sigma_d(r)$ is the dust surface density. 
This can be used to reconstruct the dust surface density as,
\begin{equation}
\Sigma_d^{\mathrm{eff}}(r)\propto \frac{\tau(r)}{2\pi r}.
\end{equation}
The overall normalization factor is proportional to the dust accretion rate ($\dot{M}_d$). For each Stokes number, we determine a single normalization factor by matching the reconstructed dust surface density to the corresponding \texttt{FARGO3D} dust surface density at the outer boundary (Section \ref{sec:code_tests}).  This provides a consistent basis for comparing the radial dust distributions obtained using the two methods across the range of Stokes numbers considered.

As a diagnostic, we also consider the theoretical radial dust drift velocity in the Nakagawa--Sekiya--Hayashi (NSH) framework \cite{nakagawa1986},
\begin{equation}
v_{r,d}^{\mathrm{NSH}} =\frac{v_{r,g}}{1 + St^2} - \frac{2\mathrm{St}\eta v_K}{1 + St^2},
\end{equation}
where $v_{r,g}$ is the radial gas velocity, $v_K$ is the Keplerian velocity, $\eta$ characterizes the departure of the gas azimuthal velocity from the Keplerian value due to radial pressure support and it is given as $\eta = - \frac{1}{2 \mathcal{M}^2} \left(\frac{d {\rm ln} P}{d {\rm ln} r}\right)$ and in our initial disk model we have $ \left(\frac{d {\rm ln} P}{d {\rm ln} r}\right) = -1$ .This expression helps us interpret $\tau(r)$, as the product $|v_{r,d}^{\mathrm{NSH}}|\tau(r)$ is expected to be approximately unity in the unperturbed disk and to deviate from unity in regions where the disk is perturbed by the planet.

Figure \ref{fig:particle_method} illustrates the particle trajectories and the corresponding radial profiles for $q=3\times10^{-4}$, the dimensionless Shakura-Sunyaev viscosity parameter,$\alpha = 0.005$, $St=0.02$, Mach Number $\mathcal{M} = 10$ and $\varepsilon = 0.5H$ where $H$ is the disk scale height and $H = \frac{r_p}{\mathcal{M}}$. In the outer disk, the trajectories are approximately circular before particles encounter the planet's perturbation. Near the planet's orbit, the trajectories deviate from this motion and establish horseshoe-like paths.

The radial profiles show how these trajectories contribute to the reconstructed distribution. The residence time decreases within the depleted region, where particles pass through more rapidly or are less frequently present, and increases in regions where particles spend longer. We apply Gaussian smoothing to the residence-time data to reduce fluctuations associated with the binning of our radial domain. The geometrical factor $1/(2\pi r)$ then converts the residence-time distribution into the effective radial surface density profile.


\section{Code Tests} \label{sec:code_tests}

Before applying the Grain Residence-Time Method to a gap-opening parameter study, we examine its numerical behavior and compare the reconstructed dust distributions with those obtained using the pressureless dust-fluid method in \texttt{FARGO3D}.

For the \texttt{FARGO3D} simulations, we set \texttt{ThicknessSmoothing = 0.5}, \texttt{FlaringIndex = 0}, \texttt{SigmaSlope = 0}, and \texttt{AspectRatio = 0.1}. The \texttt{AspectRatio} sets the reference disk aspect ratio ($h = H/r = 1/\mathcal{M}$, where $H$ is the disk scale height), while the \texttt{FlaringIndex} determines its radial dependence. Thus, our choice of zero flaring index corresponds to a constant aspect ratio throughout the disk. The \texttt{SigmaSlope} sets the power-law exponent of the initial gas surface density profile, with our choice corresponding to a radially uniform profile. The \texttt{ThicknessSmoothing} parameter sets the planetary potential smoothing length ($\varepsilon$) in units of the local disk scale height, giving a smoothing length of $0.5H$ for our adopted value. 

We neglect the back-reaction of dust on the gas. Consequently, the adopted dust-to-gas ratio only sets the normalization of the dust surface density and does not affect the gas evolution or dust dynamics. Because the dust-to-gas ratio does not impact the solution by construction and is therefore arbitrary, we choose an initial dust-to-gas ratio of unity in \texttt{FARGO3D} in order to facilitate clear comparison and code testing.  Because of the imposed scale invariance, if we were to change this dust-to-gas ratio, the result would be simple multiplication of the dust density field by an overall constant everywhere. We also assume that the planet neither accretes material nor grows during the simulations.

Unless otherwise stated, we adopt a planet-to-star mass ratio of $q=3\times10^{-4}$, a disk viscosity parameter of $\alpha=5\times10^{-3}$ near the planet (corresponding to a kinematic viscosity of $\nu=5\times10^{-5}$), and a disk Mach number of $\mathcal{M}=10$. The radial computational domain extends from $r=0.5r_p$ to $r=3.0r_p$, where $r_p$ is the planet's orbital radius.

The initial conditions for the gas and dust velocity fields and surface densities are as follows. For the gas, we adopt
\begin{equation}
v_r^g=0,\qquad
v_\phi^g=\Omega_K r\sqrt{1-h^2},\qquad
\Sigma_g=1,
\end{equation}
where $h = 1/\mathcal{M}$ is the disk aspect ratio and $\Omega_K(r)$ is the Keplerian orbital velocity. For the dust, the initial velocities are
\begin{equation}
v_r^d=0,\qquad v_\phi^d=\Omega_K r,
\end{equation}
and the initial dust surface density is given by
\begin{equation}
\Sigma_d(r)= \begin{cases} 0.01, & r<r_0,\\
                            1, & r\geq r_0,
\end{cases}
\qquad r_0=1.1 r_p
\end{equation}

By beginning with an initial cavity, we ensure that the dust must propagate completely though the system in order to achieve a steady-state.

\subsection{Time evolution of \texttt{FARGO3D}} \label{timeseries}

\begin{figure}[!t]
    \includegraphics[width=\linewidth]{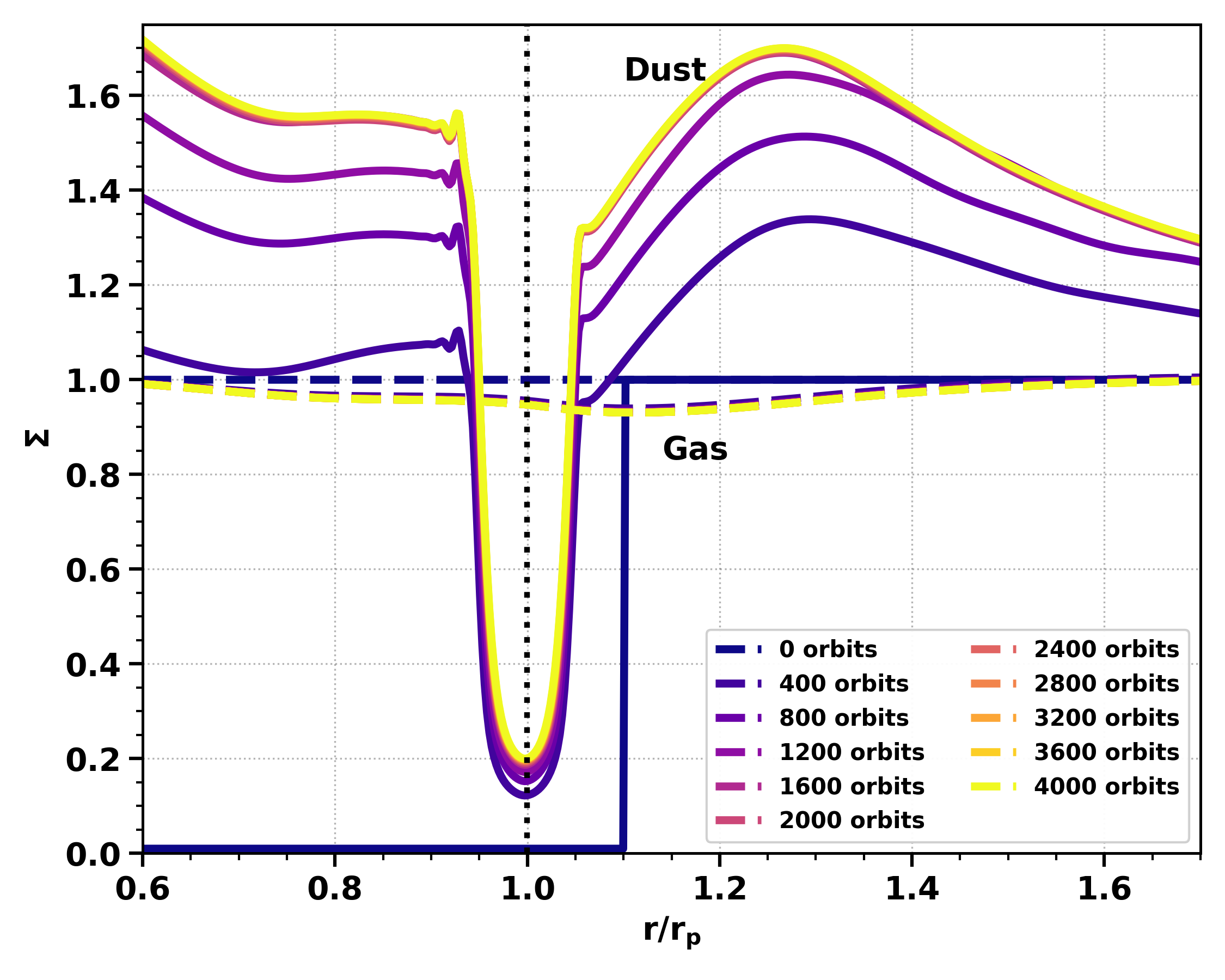}
    \caption{Azimuthally averaged gas (dashed lines) and dust (solid lines) density profiles from \texttt{FARGO3D} at different times. The parameters are $q=3\times10^{-4}$, $\alpha = 0.005$, $\mathcal{M}=10$ and $St = 0.02$.}
    \label{fig:timeseries_fargo}
\end{figure}

Figure \ref{fig:timeseries_fargo} shows the azimuthally averaged gas and dust density profiles at different times during the \texttt{FARGO3D} simulation. The gas density profile evolves relatively weakly and develops a shallow depletion near the planet, whereas the dust distribution exhibits more pronounced pile-ups and depletions.

After about 2000 orbits, the gas and dust density profiles reach a quasi-steady state. We therefore use the final output at $t=4000$ orbits as the background gas flow, with the gas velocity at each position obtained through bilinear interpolation of the velocity field.

\subsection{Convergence of FARGO3D gap profile}

\begin{figure*}[!t]
    \centering
    \includegraphics[width=0.9\linewidth]{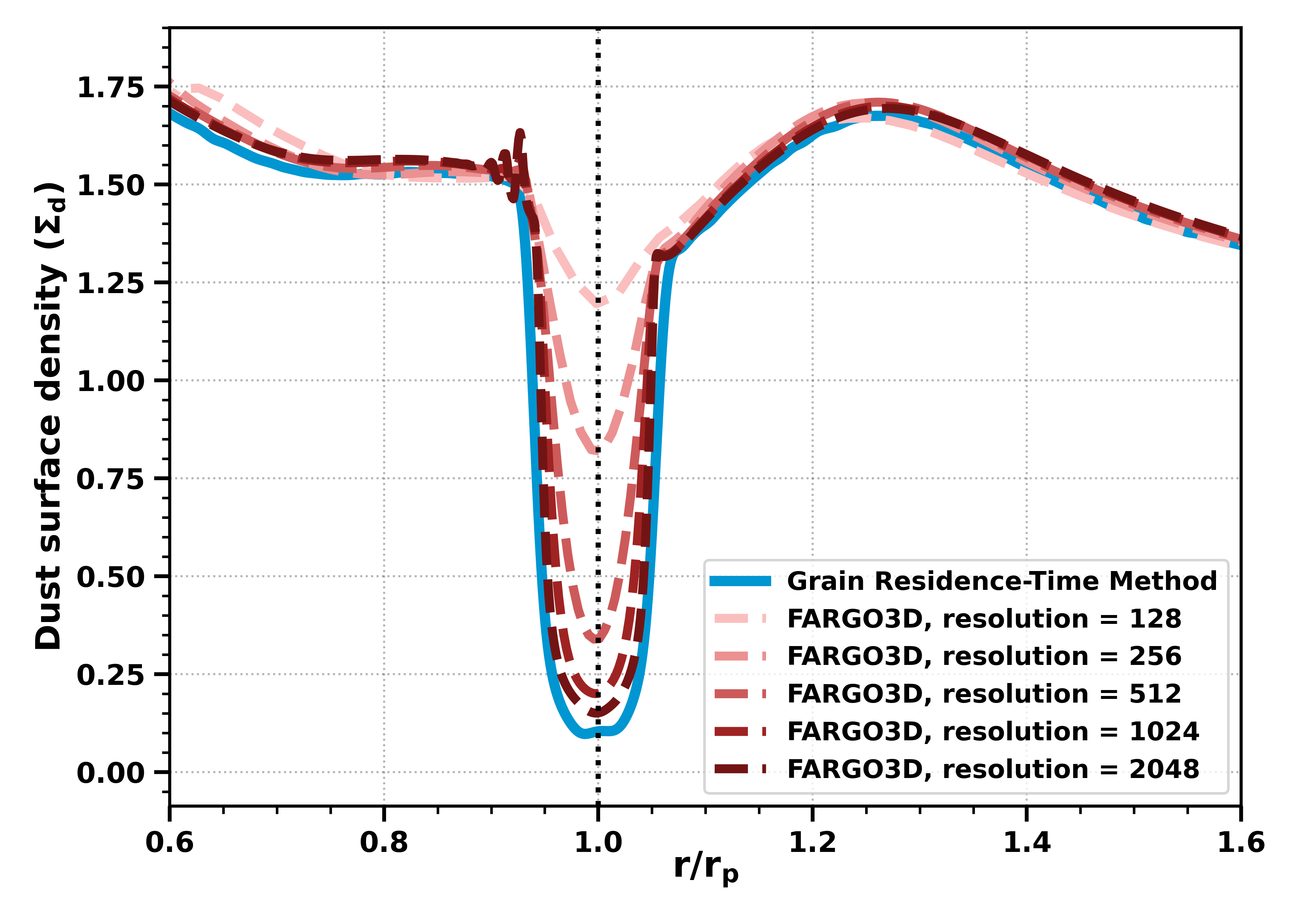}
    \caption{Azimuthally averaged dust density profiles from \texttt{FARGO3D} at different grid resolutions along with the reconstructed dust density from our Grain Residence-Time Method with five particles. The parameters are $q=3\times10^{-4}$, $\alpha = 0.005$, $\mathcal{M}=10$ and $St = 0.02$.}
    \label{fig:resolutions_fargo}
\end{figure*}

We first test convergence of the dust density profiles obtained with \texttt{FARGO3D}. Figure \ref{fig:resolutions_fargo} shows that the dust depletion becomes deeper with increasing resolution, and the $2048\times2048$ profile more closely matches the density distribution obtained with the Grain Residence-Time Method. For all the resolutions we tested, $\Delta r = \frac{2.5}{N}$, where $N$ is the resolution of the grid; our highest resolution model (N=2048) therefore has $\Delta r/r_p = 0.0012$. Outside the planet's region of influence, however, the dust density profiles remain largely similar across the resolutions tested, while differences in gap width persist.

\begin{figure}[!t]
    \includegraphics[width=0.5\textwidth]{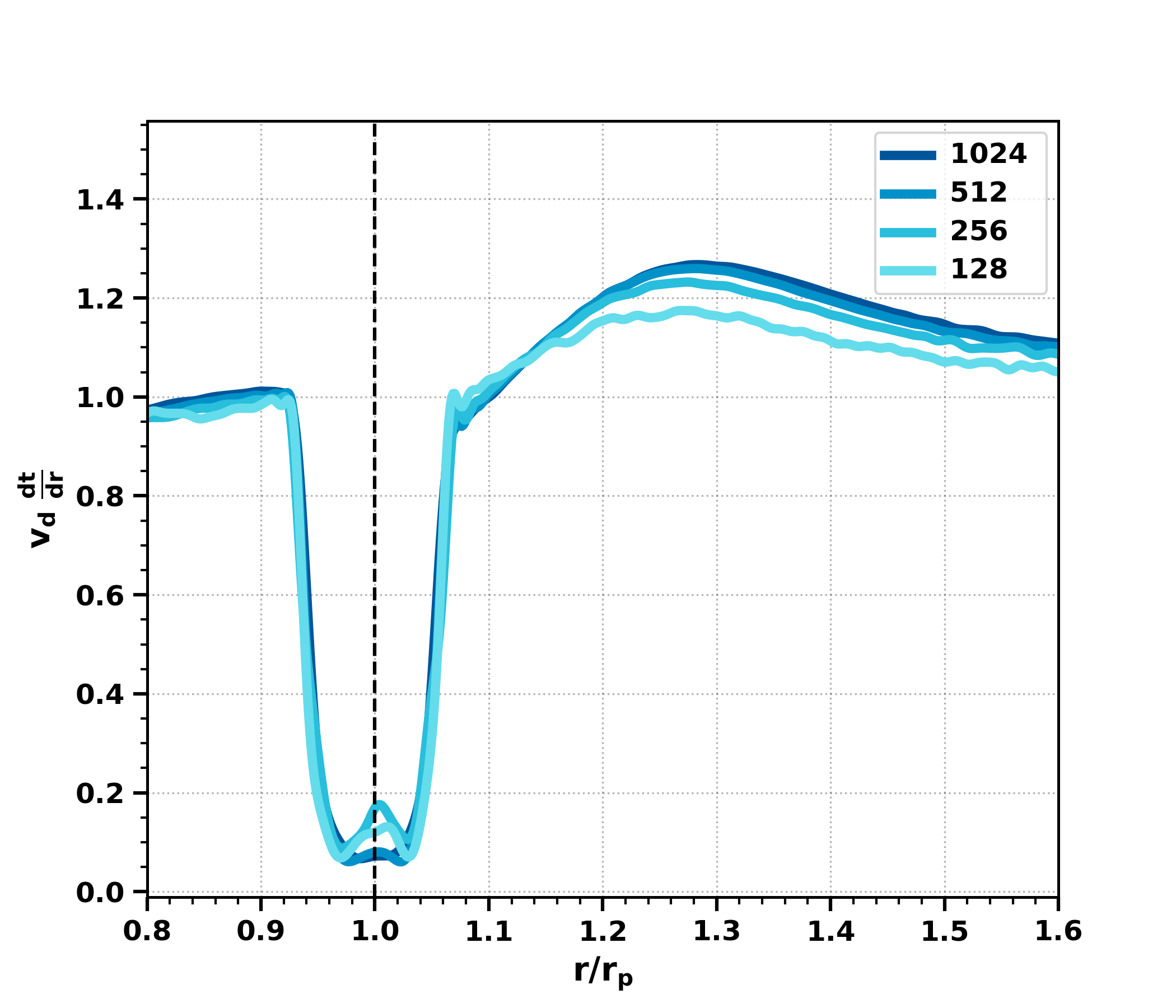}
    \caption{Reconstructed dust density profiles obtained using gas velocity fields extracted at different spatial resolutions. Five particles are used for each resolution. The y-axis shows $|v_{r,d}^{\rm NSH}|\,\tau(r)$, which is approximately unity away from the planet and deviates from unity near the planet.}
    \label{fig:res_effects_vg}
\end{figure}

We also test the effect of gas velocity field resolution on the reconstructed dust profiles. As shown in Figure \ref{fig:res_effects_vg}, the profiles remain largely unchanged across the resolutions tested, indicating that lower resolutions can be used in our parameter studies to reduce computational costs without compromising the main density features.

\subsection{Particle number convergence}

Different initial azimuthal positions expose particles to different parts of the non-axisymmetric gas velocity field, potentially producing variations in their trajectories and residence times. We therefore test how the reconstructed dust density profile depends on the number of particles used.

\begin{figure}[!t]
    \includegraphics[width=0.5\textwidth]{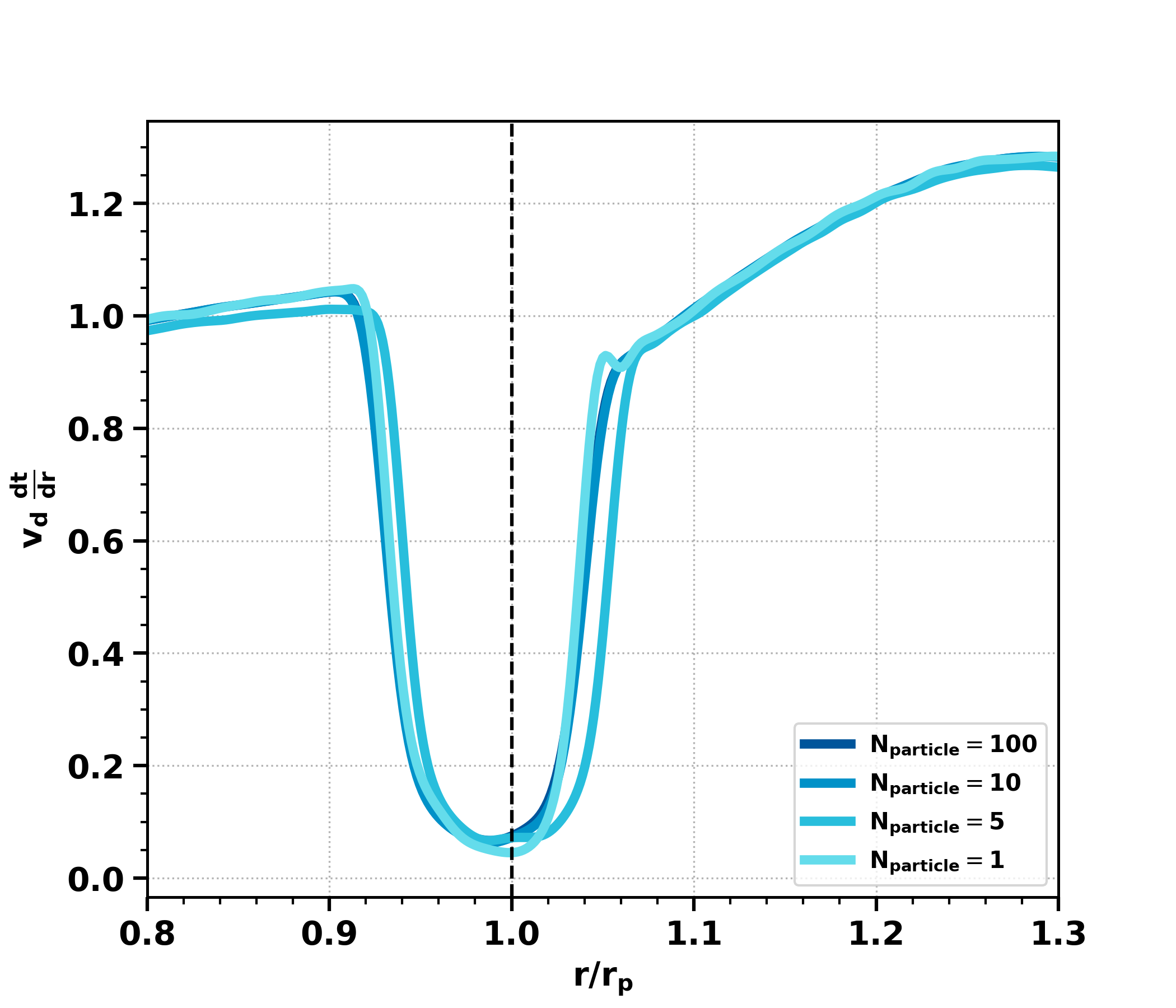}
     \caption{Reconstructed dust density profiles for different particle numbers, using the same gas velocity field with a resolution of $1024 \times 1024$. The y-axis shows $|v_{r,d}^{\rm NSH}|\,\tau(r)$, which is approximately unity away from the planet and deviates from unity near the planet.}
     \label{fig:particle_conv}
\end{figure}

We consider $N_{\rm particle}=1,\,5,\,10,$ and $100$ independent particles, initialized at the same radius and distributed uniformly in azimuth. Figure \ref{fig:particle_conv} shows that the resulting profiles are similar for all four particle numbers. For the setup considered here, a single particle therefore reproduces the main features of the reconstructed dust density distribution. This result suggests that the method can reduce the computational cost of parameter surveys, although the number of particles required may depend on the physical regime and the degree of trajectory variability. 

\subsection{Particle trajectories near the planet}

\begin{figure}[!t]
    \centering
    \includegraphics[width=0.5\textwidth]{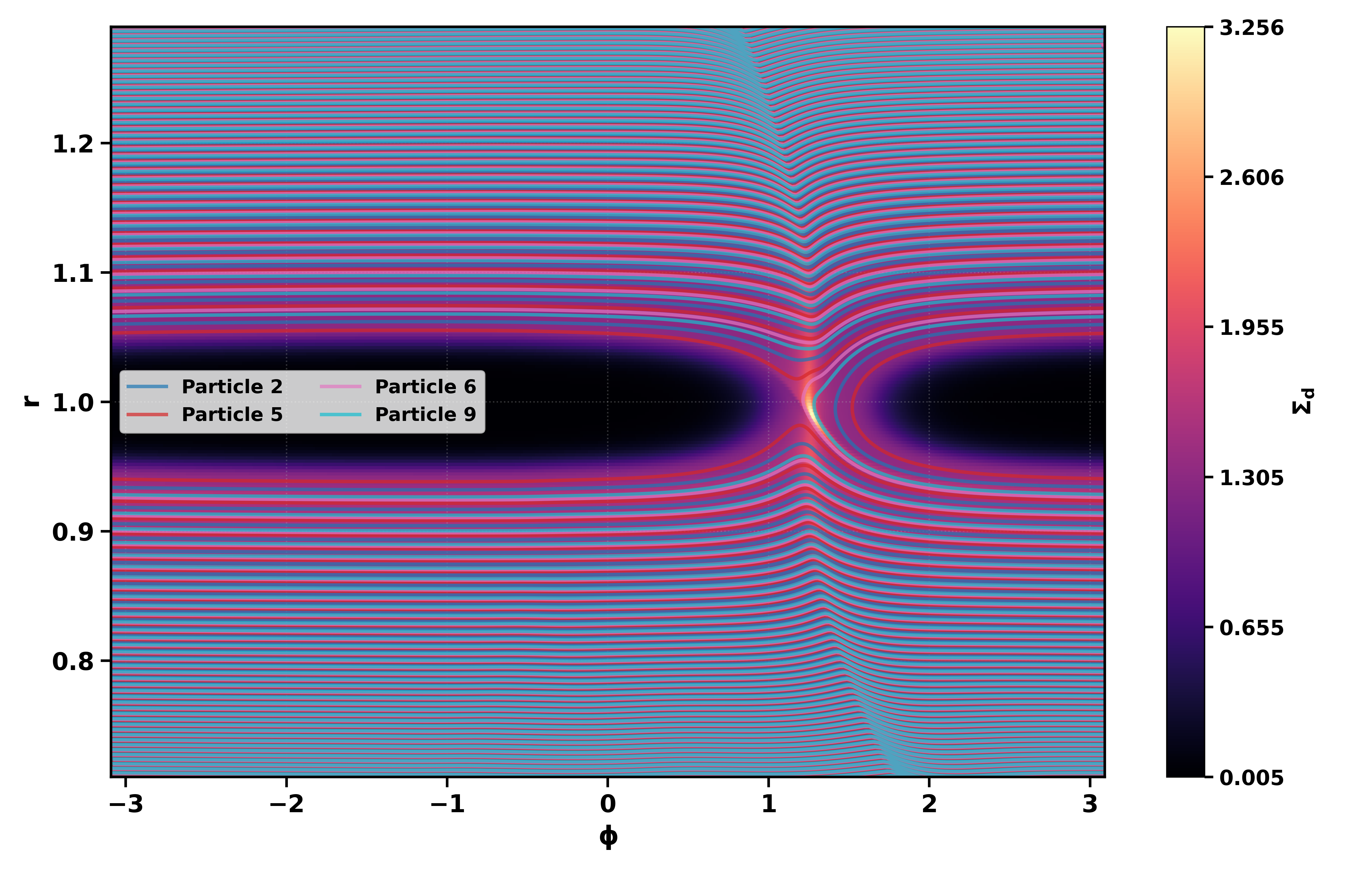}
    \caption{Selected particle trajectories overlaid on the \texttt{FARGO3D} dust density distribution.}
    \label{fig:trajectories}
\end{figure}

We examine particle trajectories near the planet to investigate how particles respond to the non-axisymmetric gas flow in the perturbed disk. Figure \ref{fig:trajectories} shows selected trajectories together with the corresponding dust density structure from \texttt{FARGO3D}. The trajectories deviate from nearly circular motion and exhibit spatial variations in the region affected by the planet.

Although the particles begin at the same radius, their different initial azimuthal positions cause them to encounter different parts of the non-axisymmetric velocity field. Their trajectories therefore differ, demonstrating that the method responds to the spatial structure of the prescribed gas flow rather than imposing purely axisymmetric radial motion.  Figure \ref{fig:trajectories} also shows that most particles preferentially pass the planet on one side when entering the inner disk. This asymmetry is also reflected in dust gap which is not symmetric with respect to the planet as depicted in Figure \ref{fig:particle_conv} for different number of particles ($N_{particle}$). 
\begin{figure}[!t]
    \centering
    \includegraphics[width=\linewidth]{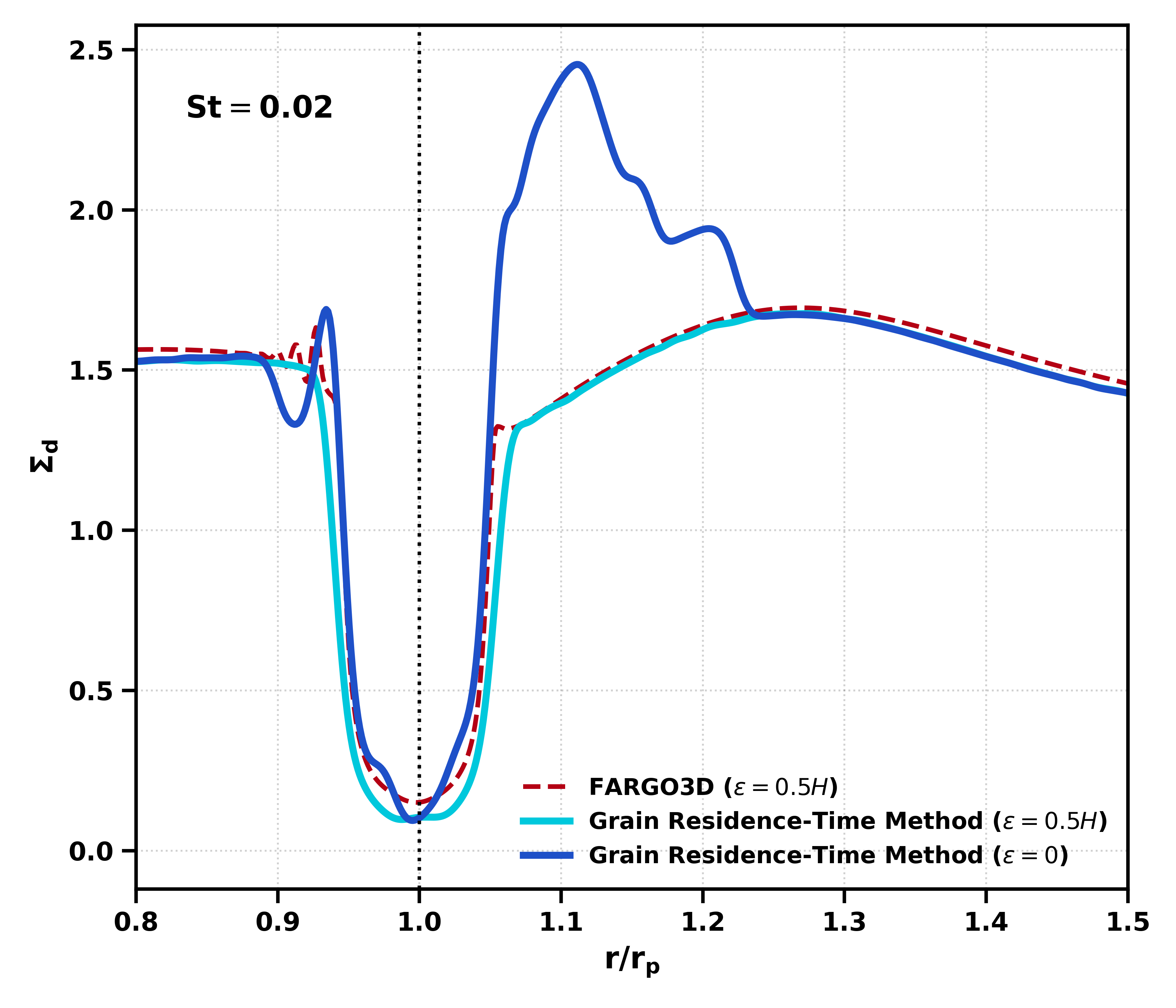}
    \caption{Comparison of reconstructed dust surface density profiles with and without planetary potential smoothing.  Because grid-based methods usually necessitate non-zero smoothing of the potential, we compare \texttt{FARGO3D} with GRTM using a smoothing length of $\varepsilon=0.5H$, where $H= \frac{r_p}{\mathcal{M}}$.  In addition, GRTM enables us to see how much the smoothing length affects the solution, in this case creating an over-density of dust on the outside of the planetary orbit when $\varepsilon = 0$.}
    \label{fig:epscomp}
\end{figure}

\begin{figure*}[!t]
    \centering
    \includegraphics[width=\textwidth]{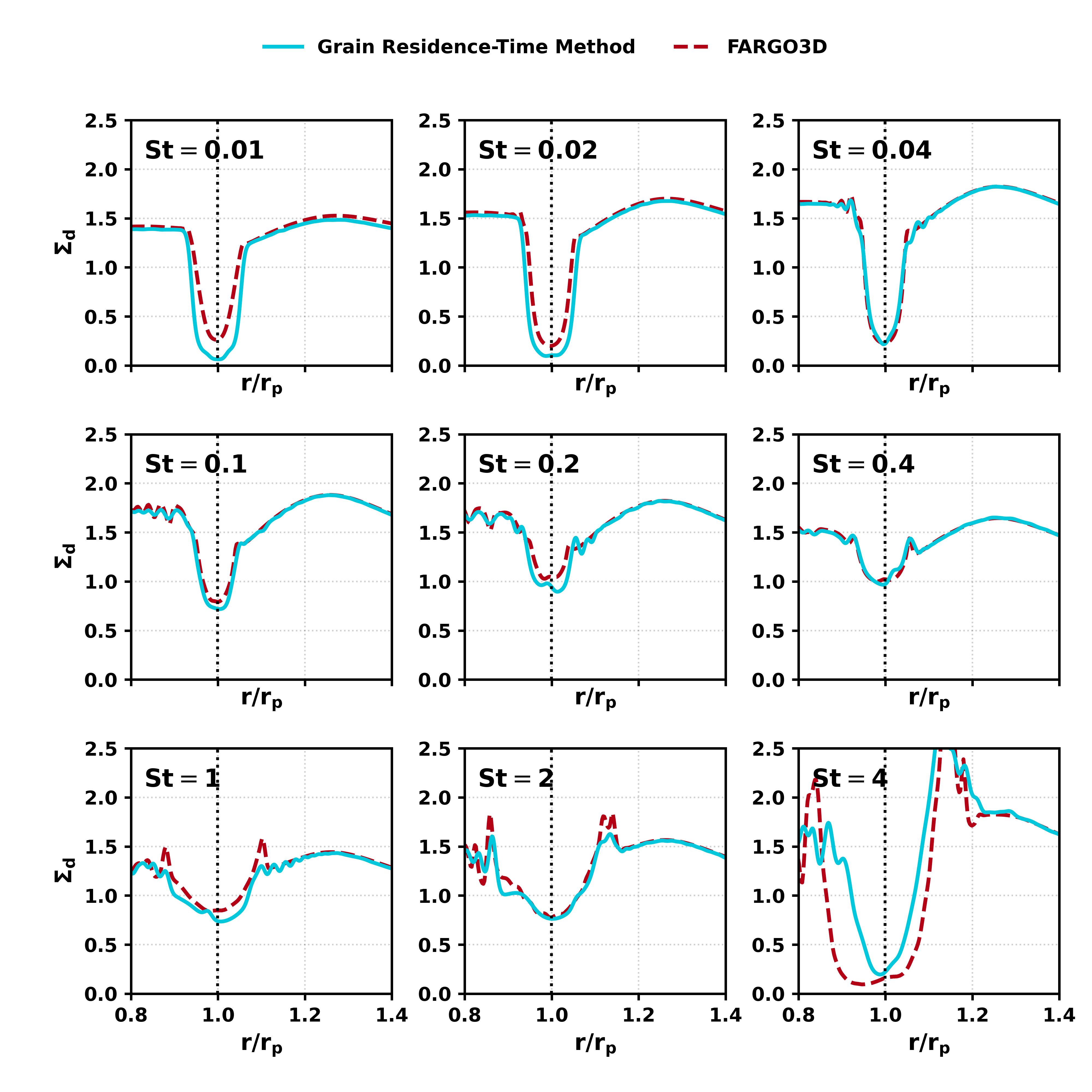}
     \caption{Steady-state dust density profiles reconstructed from particle trajectories (solid line) using the gas velocity field from \texttt{FARGO3D} at a resolution of $1024\times1024$, alongside the corresponding dust density profiles from \texttt{FARGO3D} (dashed line) at $t = 4000$ orbits, for different $\mathrm{St}$ numbers. The black dashed line shows the planet's position. All the plots have the same x and y axes limits.}
    \label{fig:smaller_sts}
\end{figure*}
\subsection{Effect of planetary potential smoothing length}
\label{subsec:eps}

In grid-based fluid simulations, the planetary gravitational potential is commonly softened to regularize the force near the
planet and, in two-dimensional models, to approximately account for the unresolved vertical structure of the disk. In
\texttt{FARGO3D}, we set \texttt{ThicknessSmoothing = 0.5}, which corresponds to $\varepsilon = 0.5H$ in our GRTM. We compare
reconstructed dust density profiles obtained with a smoothing length of $\varepsilon = 0.5H$ and with an unsoftened potential,
$\varepsilon = 0$, to assess the influence of potential softening on the dust distribution.  In both cases, however, the prescribed
gas velocity field is taken from the same \texttt{FARGO3D} simulation, which uses a softened planetary potential with $\varepsilon = 0.5H$.
Thus, only the softening length in the direct gravitational force acting on the particles is changed.

Figure \ref{fig:epscomp} shows that, for the parameters considered, the unsoftened potential produces an enhanced dust
density exterior to the planet's orbit. This occurs because particles passing near the planet are deflected outward to larger radii and subsequently move back inward, increasing their residence time in this region. Across the parameter space explored, particle trapping near the planet's orbital radius is also frequently observed when potential softening is enabled. The ability to use an unsoftened planetary
potential ($\varepsilon=0$) in the GRTM therefore allows us to investigate how the direct gravitational influence of the planet affects particle trajectories and the resulting dust pile-ups near the outer edge of the gap. 

\subsection{Dependence on Stokes number}

We test the method over a range of Stokes numbers, from $St=0.01$ to $St=40$, spanning particles that are strongly to weakly coupled to the gas. However, we do not expect the pressureless dust-fluid method to be as accurate at higher Stokes numbers, as it is most applicable when particles are at least moderately coupled to the gas. Therefore, for high $\mathrm{St}$ numbers this is a test for whether the fluid approximation breaks down.

\subsubsection{Lower Stokes numbers: $St < 1$}

For $St<1$, the particle and fluid methods produce broadly similar radial dust density profiles (Figure \ref{fig:smaller_sts}). The agreement is particularly good away from the planet, where the planetary perturbation is weaker. Both methods also reproduce similar dust depletion features near the planet, indicating that the particle-based reconstruction captures the main radial structures in this regime.

\begin{figure}[!t]
    \centering
    \includegraphics[width=0.45\textwidth]{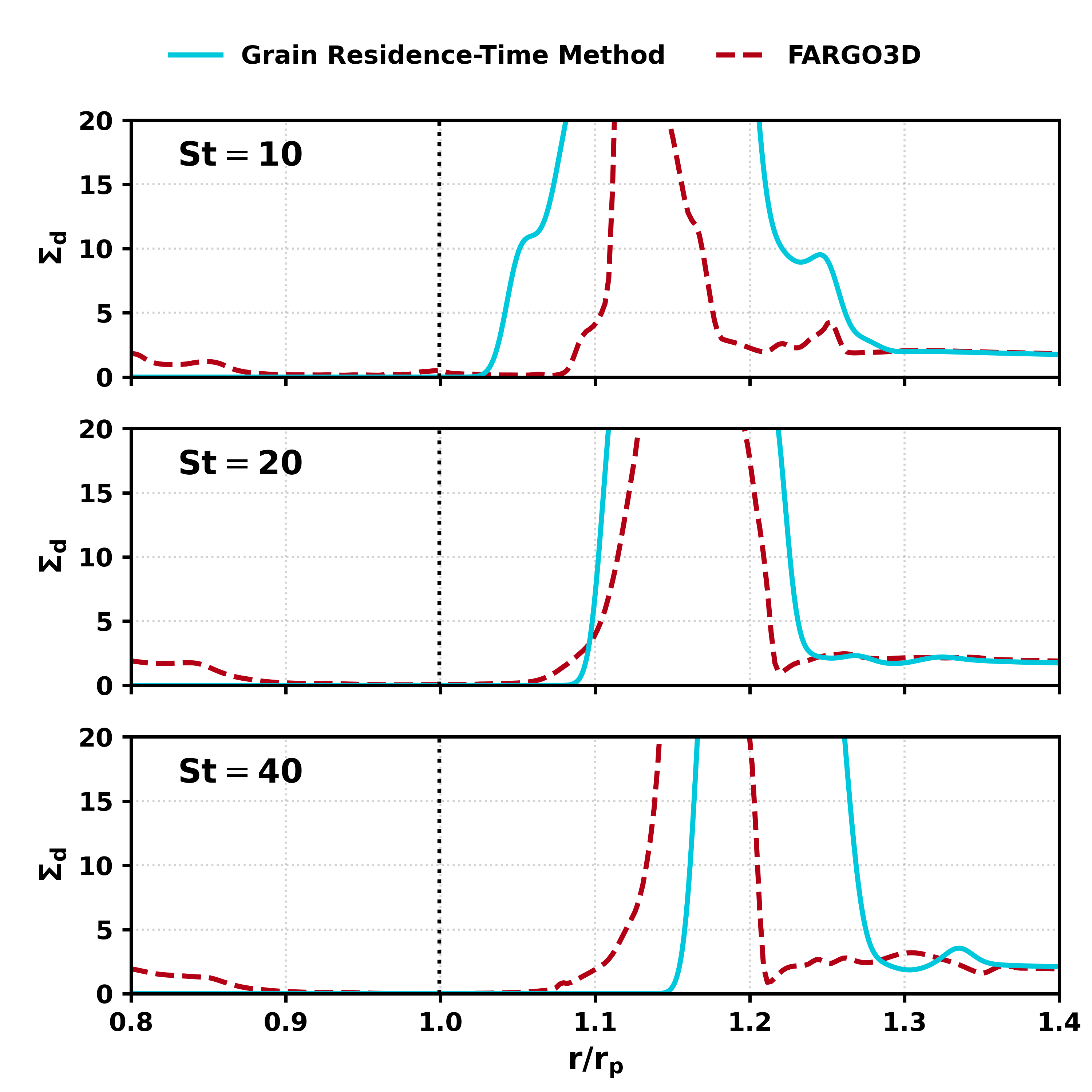}
    \caption{Dust density profiles reconstructed from particle trajectories using the gas velocity field from \texttt{FARGO3D} at a resolution of $1024\times1024$, alongside the corresponding dust density profiles from \texttt{FARGO3D} at $t = 4000$ orbits, for $\mathrm{St}$ numbers where the particles are trapped outside the planet. The black dashed line shows the planet's position.}
    \label{fig:higher_sts}
\end{figure}
\subsubsection{Higher Stokes numbers: $St > 1$}

For $1<St<10$, the two methods continue to show some common qualitative features, including dust pile-ups and depletions, but differences between their detailed radial profiles become more pronounced (Figure \ref{fig:smaller_sts}). The \texttt{FARGO3D} dust profiles also become increasingly oscillatory as $\mathrm{St}$ increases. At higher Stokes numbers (Figure \ref{fig:higher_sts}), aerodynamic drag is less effective at coupling the particle motion to the gas, and the planetary gravitational perturbation plays a more important role in determining the trajectories.

For $St\geq10$, particles in our method accumulate near resonant radii and do not cross the planet's orbital radius during the integration. Such parameters are marked as ``trapped outside" in Section  \ref{sec:results}. And in such cases, the density peaks become more pronounced with integration time, consistent with particles spending extended periods near these locations. The \texttt{FARGO3D} profiles also exhibit pile-ups near resonant locations; however, an azimuthally averaged density profile alone cannot establish whether individual particles remain trapped or cross the planet's orbit.

The locations of the density peaks vary with $\mathrm{St}$, consistent with the locations of different mean-motion resonances. These results illustrate an advantage of the residence time approach: tracking individual trajectories
provides direct information about particle crossing and trapping behavior that cannot be inferred from an azimuthally averaged fluid density profile alone.

\section{Parameter Survey} \label{sec:results} 
Using the Grain Residence-Time Method, we perform a parameter survey over the planet-to-star mass ratio ($q$) and Stokes number ($\mathrm{St}$) to investigate how dust gap formation depends on planetary mass and particle-gas coupling. Figure \ref{fig:param_survey} shows the gap diagnostic $G$ across the parameter space, where $G$ measures the strength of the dust density depletion around the planet's orbital radius.
\begin{align}
    G &= 1 - \frac{\Sigma_d}{\Sigma_0} \\
    \frac{\Sigma_d}{\Sigma_0} &=|v_{r,d}^{NSH}| \times \tau(r)   
\end{align} 
With larger values of $G$ corresponding to a deeper gap. Since we run multiple particles with different initial azimuth, we consider the average value of $G$, which is shown as $G_{av}$ in Figure \ref{fig:param_survey}.

We identify four characteristic regimes based on the particle trajectories and reconstructed dust density:

\begin{figure*}[!t] 
    \hspace{1cm}
    \includegraphics[width=0.8\textwidth]{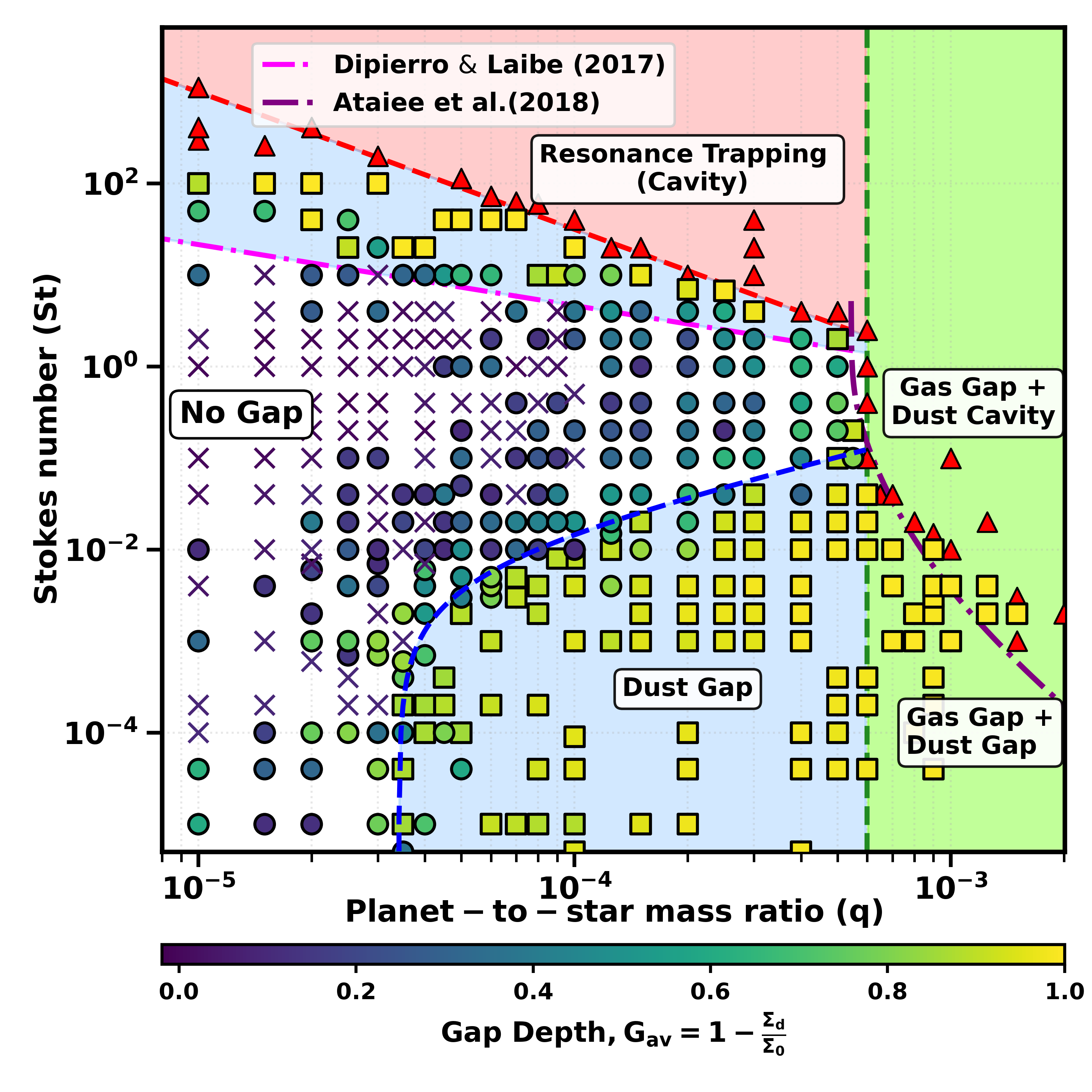}
\caption{Parameter survey of the reconstructed dust density profiles as a function of planet-to-star mass ratio $q$ and Stokes number $\mathrm{St}$. The profiles are classified into four characteristic regimes: no gap (cross), partial gap (circle), gap (square), and trapped outside (triangle). The color bar shows the averaged gap depth. We considered $\varepsilon = 0$ for this parameter survey.} 
\label{fig:param_survey}
\end{figure*}

\begin{itemize}
\item \emph{No gap}: Particles cross the planet's orbit with little deviation from their inward radial drift. $G_{\mathrm{av}} < 0.1$.

\item \emph{Partial gap}: Particles cross the planet's orbit but, rather than being completely deflected by the planet, spend some time near it, resulting in a partial gap. $0.1 < G_{\mathrm{av}} < 0.85$.

\item \emph{Gap}: Particles cross the planet's orbit, but their trajectories are strongly perturbed by the planet, causing them to avoid the planet's Hill sphere and make little to no significant contribution to the local reconstructed dust density. $G_{\mathrm{av}} > 0.85$.

\item \emph{Trapped outside}: Particles remain outside the planet's orbit and accumulate at locations exterior to it. In this case, $G_{\mathrm{av}}$ can also approach 1, so the the inner disk density is used to distinguish this regime from a gap.
\end{itemize}

The classification therefore uses both the average gap diagnostic, $G_{\mathrm{av}}$, and the inner disk dust density to distinguish particles that cross the planet's orbit from those that remain trapped outside it.

\subsubsection{Dust gap opening in the lower $\mathrm{St}$ regime} \label{lowstgap}

An interesting result of our parameter survey is the formation of dust gaps at low $\mathrm{St}$ numbers, which generally correspond to smaller grains for a given local gas environment. These particles are more strongly coupled to the gas, making the formation of significant dust depletion particularly surprising, especially since the corresponding gas gaps are relatively shallow. This low $\mathrm{St}$ regime may therefore be relevant to interpreting dust substructures observed with ALMA, although we do not make a direct comparison with observations in this work.

To better understand the particle dynamics associated with these gaps, we examine their trajectories near the planet. As shown in Figure \ref{fig:trajectories}, particles passing near the planet are deflected, with most passing on its leading side before continuing toward the inner disk. The reconstructed radial dust surface density profiles also reveal an asymmetry in the gaps, with the inner edge lying farther from the planet's orbital radius ($r_p$) than the outer edge. We find that the distance from $r_p$ to the outer edge of the gap is comparable to the planet's Hill radius, $R_H = r_p(q/3)^{1/3}$. \cite{dipnlaibe} found a similar characteristic scale in their higher-$\mathrm{St}$ regime, with the distance from the planet's orbit to the outer edge of the dust gap ($\Delta_{\rm gap}$) bounded by $R_H \lesssim \Delta_{\rm gap} \lesssim H$. However, we have not yet carried out a systematic measurement of the gap widths across our parameter space, so this comparison remains qualitative.

\subsection{The gap-opening criterion} \label{gap opening}
The different dust-density structures occupy distinct regions of the $(q,St)$ parameter space. We identify empirical scalings that approximately separate these regions, as shown in Figure \ref{fig:param_survey}.

\begin{itemize}
 \item For higher $\mathrm{St}$ numbers, the particles are trapped for $St \gtrsim St_{\rm Trap} =  \left( q/q_{0}  \right) ^{-1.5} $ with $q_{0} = 0.001$. This scaling was found empirically. So we do not know how $q_{0,trap}$ scales with $\mathcal{M}$ and $\alpha$. However, in the range $10^{-2} <\mathrm{St}< 10^2$, the grains drift much faster than the viscous drift rate and therefore it is expected this scaling does not depend on $\alpha$. One possibility is that $q_{0,trap}$ is given by the transition to nonlinearity in which case $q_{0,trap} \approx \mathcal{M}^{-3}$.
 
 \item For smaller $\mathrm{St}$ values ($\mathrm{St} < 1 $) we obtain a semi-empirical criterion;  $St \lesssim St_{\rm gap} = \mathcal{M}^2\ (2.2\ q- 1.5 \alpha / \mathcal{M}^2 ) $, which follows from requiring the dust drift rate, $ v_{d}/v_k $ to be less than $2.2 \ q$, where $v_k = \Omega_K r$. The scaling $2.2 q$ is an entirely empirical scaling and we will try to explore this more in our future work.
\end{itemize}
We next compare these empirical scalings with previously proposed criteria and discuss where they compare and break down.

The region highlighted in red and labeled as Resonance Trapping (cavity), is bounded by our empirical scaling from below. Related resonance trapping has been studied in previous
work, which established a critical particle size for
passage through a given resonance \citep{weidenschillingetal}. Expressed in terms of the
Stokes number, the criterion is
\begin{equation}
    St_{min} = \frac{\sqrt{\frac{\pi}{8}}}{3 q C(j) j^{3/2}}
\end{equation} 
For large $j$, the outer $(j+1):j$ resonance locations satisfy $(r_{\rm res}-r_p)/r_p \simeq 2/(3j)$. Combining this with the resonance separation scaling of \cite{Paardekooper_dustpart}, $(r_{\rm res}-r_p)/r_p \propto q^{2/7}$, gives $j\propto q^{-2/7}$. Assuming the Laplace coefficient at $j(j+1)$ resonances, $C(j)\propto j$, which is reasonably accurate for $j \sim 10$, the criterion gives  $St_{\min}\propto q^{-2/7}$. This dependence is significantly shallower than our empirical scaling and does not match the parameter-space region in which particles become trapped in our simulations.

We also see dust gaps in two different $\mathrm{St}$ number region (shaded as blue in Figure \ref{fig:param_survey}).  \cite{dipnlaibe} derived the criterion to have dust gaps as $St \approx \mathcal{M}^{-2} q^{-2/3}$, which is shown in Figure \ref{fig:param_survey}.
For smaller mass ratios, where planets carve a shallow gap in gas,we are able to get full gaps in dust for different $\mathrm{St}$ numbers according to our empirical scaling law. This result is consistent with the conclusion of \cite{dipi_gas_dust} that a gas gap is not a necessary condition for dust-gap formation around low-mass planets.

However, both empirical scalings break down as \(q\) increases, particularly for \(q \gtrsim 6\times10^{-4}\), where the region above this threshold is shaded green in Figure \ref{fig:param_survey}. This behavior is expected because for sufficiently high mass, the planet opens a gap in the gas, which is expected to lead to either a gap or cavity in the dust. \cite{duffelgasgap} found a gas-gap opening criterion with the scaling $q_{\mathrm{gap}} \propto \sqrt{\alpha}\mathcal{M}^{-5/2}$.  These gaps modify the gas pressure gradient, producing pressure maxima where dust particles can become trapped.  Thus, our empirical criterion for the green shaded region is

\begin{equation}
    q > q_{\rm gas} = 2.7 \alpha^{1/2} \mathcal{M}^{-5/2}.
\end{equation}

The criterion for dust trapping in presence of deep gas gap has been well-studied by \cite{ataiee} and their semi-analytical criterion can be rewritten as,
\begin{equation}
   \mathrm{St}= \frac
    {1}{ \sqrt{ 99.25 \frac{h^2}{\alpha}\left(\frac{q}{q_{g,A}}-1.\right) ^{\frac{20}{7}}-4.0}}
\end{equation}
Where, $q_{\mathrm{g,A}} = h^3 \sqrt{37.3\alpha + 0.01}$ is their planet-to-star mass ratio above which a radial pressure maximum develops beyond the planet's orbit. As shown in Figure \ref{fig:param_survey}, this criterion is reasonably consistent with the trapping behavior in our parameter-space results for $q \gtrsim 6\times10^{-4}$.

Dust grains at low $\mathrm{St}$ values are expected to closely follow the gas and drift slowly relative to it. We therefore anticipate that, at sufficiently low $\mathrm{St}$, particles may spend more time in the vicinity of the planet, allowing planetary perturbations to act on them for longer and potentially compete with viscous effects. This may lead to a broader region in which no dust gap forms. However, no such region was found within the parameter space explored in this study, and it may emerge at even lower $\mathrm{St}$ values.

\section{Discussion}
\label{sec:discussion}

We introduce the Grain Residence-Time Method (GRTM) as an efficient approach for studying dust density features in protoplanetary disks. Using a prescribed gas velocity field from \texttt{FARGO3D}, we reconstruct the radial dust distribution from particle residence time and compare it with the surface dust density obtained from \texttt{FARGO3D} using the pressureless dust-fluid approach. We find good agreement between the two methods for most of the lower $\mathrm{St}$ numbers considered in this study. At higher $\mathrm{St}$, the methods begin to differ, particularly near resonant peaks, where the dust-fluid approach produces broader and smoother pile-ups than the particle-based method. These differences may partly reflect limitations of the pressureless-fluid approximation at higher Stokes numbers. A pressureless-fluid model assumes a single dust velocity at each spatial location, whereas particles following different trajectories can pass through the same region with different velocities. As $\mathrm{St}$ increases, aerodynamic drag becomes less effective at damping these velocity differences, making the single-velocity approximation less well justified \citep{jacquet2011}. The resolution dependence of the \texttt{FARGO3D} dust profiles, discussed in Section \ref{sec:code_tests}, may also contribute to these differences.

By exploring a wide range of planet-to-star mass ratio values ($q$) and Stokes numbers ($\mathrm{St}$), we identify two major empirical scaling relations and regimes in the parameter survey. (i) $St \propto q^{-1.5}$ , which separates the region of possible full gaps as previously predicted \citep{dipnlaibe} and identify regions where particles become trapped outside the planet's orbit, producing a dust pile-up, and (ii) $St \propto \mathcal{M}^2 q$ which distinguish full gaps from partial gaps in the lower $\mathrm{St}$ regime. 
We also report proportionality constants of these scalings in Section \ref{gap opening}. However, we considered a single combination of Mach number ($\mathcal{M} = 10$) and viscosity parameter ($\alpha = 0.005$) in all of our studies presented here. Hence, we cannot determine how the scaling relations would change with these parameters. 
Since we classify the reconstructed dust profiles based on the criteria we set on the average gap-depth, $G_{\mathrm{av}}$, the coefficients on these scalings could possibly have an additional dependence on this criteria as well. Along with exploring different disk parameters and gap depth criteria, future works will also investigate how the widths and depths of dust gaps depend on $\mathrm{St}$, $q$, $\alpha$, and $\mathcal{M}$. This would possibly help us understand the asymmetry of the radial dust density gaps with respect to the planet that we noted in Section \ref{lowstgap}.

For smaller $\mathrm{St}$, we obtain a semi-empirical criterion for dust-gap opening by requiring the dust drift rate to satisfy $v_d/v_K \lesssim 2.2q$. As discussed in Section \ref{gap opening}, this gives a boundary of the form $St \lesssim \mathcal{M}^2(2.2q-1.5\alpha/\mathcal{M}^2)$. While this criterion captures the boundary in the low $\mathrm{St}$ regime accurately, it predicts that toward low $q$ end of the parameter space, the boundary approaches an almost constant value of $q$. For our adopted $\alpha$, this limiting value is $q \approx 3.5 \times 10^{-5}$, resulting in a nearly vertical boundary in the parameter space plot (Figure \ref{fig:param_survey}). Since our survey extends only down to $St \gtrsim 10^{-5}$, extending it to smaller $\mathrm{St}$ numbers will help determine whether this trend continues. Future work will also investigate the predicted dependence of this boundary on $\alpha$.

Both empirical scalings also break down at higher planet-to-star mass ratios, particularly for $q \gtrsim 6\times10^{-4}$. In this regime, the planet-induced gas structure becomes increasingly important, as the gas gap and associated pressure maximum can alter the radial drift of dust particles and lead to trapping outside the planet's orbit. This behavior is distinct from the resonant trapping identified at higher $\mathrm{St}$ and is consistent with the pressure-maximum criterion of \cite{ataiee}, which reasonably reproduces the trapping region in our parameter survey at these mass ratios.

As pointed out by \cite{Price_2025}, assuming a constant physical grain size is more appropriate than assuming a constant $\mathrm{St}$ throughout the disk, because the gas density can change substantially near the planet. In the Epstein regime, $\mathrm{St}\propto \rho_\bullet a/\Sigma_g$, where $\rho_\bullet$ and $a$ are the grain internal density and physical size, respectively. Thus, a fixed physical grain size generally corresponds to a spatially varying Stokes number when the gas surface density varies. In future work, we will use physical grain sizes and calculate the corresponding $\mathrm{St}$ number from the local gas conditions, allowing the aerodynamic coupling to vary throughout the disk. Currently GRTM does not account for dust growth or interactions between dust grains. Including these processes would allow us to study a broader range of dust evolution in protoplanetary disks.

The particle dynamics also depend on the gas velocity field interpolated from \texttt{FARGO3D} outputs. Interpolation errors may affect the reconstructed dust distribution, particularly close to the planet, where the gas flow is strongly perturbed. In future work, we plan to explore different approaches to developing a semi-analytical prescription for the local gas velocity, which could reduce these errors and make the method less dependent on the simulated velocity field.

Since the GRTM follows individual particle trajectories, it can capture the dynamics of particles at higher $\mathrm{St}$, including resonant trapping. In future work, we will build on this to study dust filtration and quantify how much dust is prevented from crossing the planet's orbit. We also plan to post-process the reconstructed dust distributions with radiative transfer, allowing us to compare the dust structures predicted by our method with observations and assess the relevance of our criteria to observed protoplanetary disks.

Our code tests (Section \ref{sec:code_tests}) show that the reconstructed dust profiles remain largely unchanged when the resolution of the prescribed gas velocity field is varied. The particle-number test further shows that a single particle can reproduce the main density features for the setup considered, although the number of trajectories required may vary across different parameter regimes. In contrast, the \texttt{FARGO3D} dust-fluid profiles show a stronger dependence on spatial resolution, with deeper dust depletion at higher resolutions. Since the GRTM does not require evolving the dust as a fluid on a grid, it provides a computationally efficient approach for carrying out large parameter surveys of dust dynamics and gap formation.


    

\begin{center}
\textbf{Acknowledgments}
\end{center}
\vspace{-5pt}
\noindent
We would like to thank Levi Walls, Eugene Chiang and Ellen M. Price for their helpful comments and suggestions. The \texttt{FARGO3D} simulations were performed on the Petunia Cluster, hosted by the Department of Physics and Astronomy. We also thank Chris J. Orr for maintaining the cluster.


\bibliography{sample701}{}
\bibliographystyle{aasjournal}

\end{document}